# Pressure-induced enhancement of two-dimensionality in LaO$_{1-x}$F$_x$Bi(Se/S)$_2$ superconductors

V. Svitlyk*[1], A. Krzton-Maziopa[2], M. Mezouar[1]

[1]ID27 High Pressure Beamline, European Synchrotron Radiation Facility, 38000 Grenoble, France

[2]Faculty of Chemistry, Warsaw University of Technology, Noakowskiego 3, 00-664 Warsaw, Poland

*svitlyk@esrf.fr

**Abstract**

Application of high pressure (HP) induces evolution from metallic to semi-metallic state in the layered superconducting LaO$_{1-x}$F$_x$BiSe$_2$ phase, as concluded from *ab initio* calculations based on experimental *P*-dependent structural data. These changes in conduction properties are associated with a reported decrease in the superconducting transition temperature, $T_c$. However, further increase in pressure induces structural transformation in LaO$_{1-x}$F$_x$BiSe$_2$ which results in a strong enhancement of its two-dimensionality with a related improvement in superconducting performance. In addition, the HP structural anisotropy induces a loss of long-range order correlations along the stacking direction of the LaO/F and BiSe$_2$ layers yielding formation of stacking faults. Similar but even more pronounced structural transition was observed in the related LaO$_{1-x}$F$_x$BiS$_2$ phase which is also accompanied by an increase in $T_c$. This demonstrates that enhancement of two-dimensionality in layered superconducting systems acts in favor of their superconducting performance.

**Introduction**

Layered 2D systems form a major class of superconductive materials with classics like Cu-oxide and Fe-based compounds. Yet another family of layered BiCh$_2$-based superconductors (Ch – chalcogenide S or/and Se atoms) have been discovered afterwards (Bi$_4$O$_4$S$_3$ [1,2]) thus further enriching this vast constellation of 2D superconductive systems. The BiCh$_2$-based materials have a typical for layered superconductors structure. Namely, they are composed of conducting BiCh$_2$ layers which are separated by blocking (spacer) layers which serve as charge reservoirs (Fig. 1, LaO$_{1-x}$F$_x$BiSe$_2$ phase is shown). This modular structural arrangement allowed synthesis of numerous derivatives of the parent Bi$_4$O$_4$S$_3$ phase [3], similarly to Cu-oxide and Fe-based systems.

Application of external pressure on the BiCh$_2$-based compounds substantially alters their superconducting response. For instance, the LaO$_{0.5}$F$_{0.5}$BiSe$_2$ phase with a superconducting transition temperature of about 3 K [4,5] features a decrease of $T_c$ as a function of pressure [5,6] (we denote this phase as SCI, *P*4/*nmm* symmetry). However, upon a further compression a new phase with an increased $T_c$ ~ 6 K emerges (SCII phase) [5-7]. Emergence of the HP SCII phase was observed in many other BiCh$_2$-based materials [8-17], including LaO$_{0.5}$F$_{0.5}$BiS$_2$ [17-20].

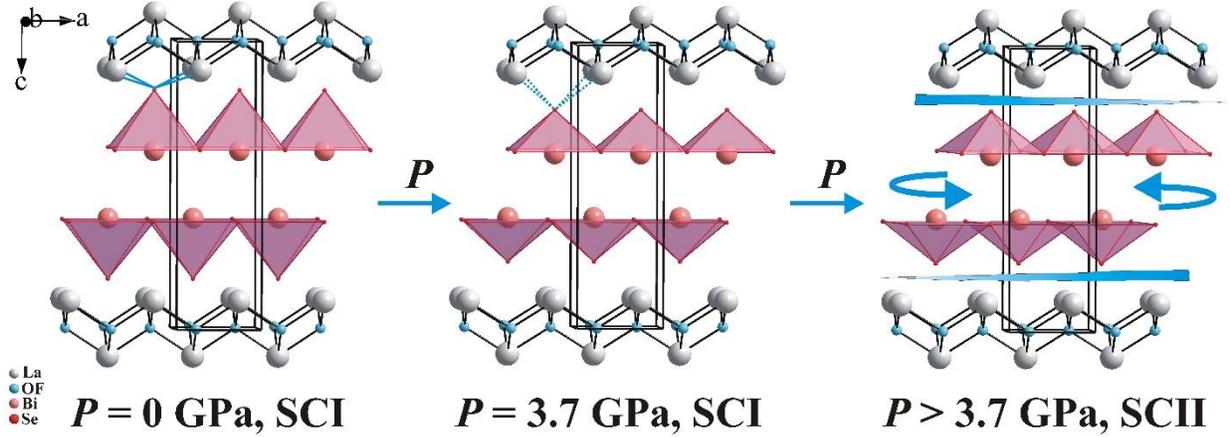

Fig. 1. *P*-induced evolution of the SCI phase and formation of the SCII phase of LaO$_{1-x}$F$_x$BiSe$_2$. Rotation arrows (right) schematically indicate decoupling between the BiSe$_2$ bilayers and the reference LaO$_{1-x}$F$_x$ framework resulting in *P*-induced stacking faults.

The SCII phase was reported to crystallize in a monoclinic structure ($P2_1/m$), as concluded from *P*-dependent powder diffraction data [19]. The proposed HP model for the SCII phase does not, however, satisfactorily fit experimental data ($R_{wp}$ = 25,5%) [19]. Interestingly, for certain BiCh$_2$-based phases superconducting response at ambient pressure can be enhanced in a similar way as for SCII phases by using a HP annealing technique [21-25] and this phenomena was reported to be related to the induced micro-structural changes along the *c* axis, in particular manifested through a peak broadening. Similar broadening was observed for the SCII phase [19]. Clearly, detailed *P*-dependent structural studies on BiCh$_2$-based materials are needed to reveal formation mechanism and nature of the SCII phase in order to better understand the associated improvement in superconducting performance. Here we report single crystal synchrotron radiation diffraction studies on LaO$_{1-x}$F$_x$BiSe$_2$ and LaO$_{1-x}$F$_x$BiS$_2$ compounds as a function of external pressure and temperature. Related *ab initio* calculations have been performed to correlate experimental structural behavior and observed superconducting response at HP.

**Experimental details**

The LaO$_{1-x}$F$_x$BiSe$_2$ and LaO$_{1-x}$F$_x$BiS$_2$ monocrystalline samples have been grown via high temperature flux method using high purity (at least 99.99%, Alfa) powders of La$_2$O$_3$, LaF$_3$ and Se/S, and metallic La. For crystal growth 1g of raw materials, weighted in appropriate molar ratio [4] was blended with 5 g of flux composed of RbCl and NaCl (mixed in molar ratio: 0.56:0.44) and sealed in a double wall evacuated quartz ampoules. Ampoules were heated at 800°C over 15 h and then slowly (2 deg/hour) cooled down to 550°C, and then fast cooled to room temperature. Next the ampoules were opened in air and crystals were extracted by dissolving chloride flux in distilled water. Afterwards crystals were washed repeatedly with acetone and dried in air. The exact chemical compositions, as obtained from EDS analysis, are LaO$_{0.76(1)}$F$_{0.24(1)}$BiSe$_2$ and LaO$_{0.68(2)}$F$_{0.32(2)}$BiS$_2$ with corresponding $T_c$ at ambient pressure of 2.6 and 3 K, respectively.

*P*- and *T*-dependent synchrotron radiation diffraction data were collected at the ID27 High Pressure Beamline, European Synchrotron Radiation Facility, Grenoble, France. Initially, LaO$_{1-x}$F$_x$BiSe$_2$ was compressed at room temperature (RT) up to 55 GPa with a typical step of 2 GPa. For this the crystal was contained in a rhenium (Re) gasket with a hole of 125 µm ($D_g$) mounted on a diamond anvil cell (DAC) with 250 µm culets ($D_c$). The data were recorded on a flat panel Perkin Elmer detector. In order to study the SCI-SCII transition in greater details the LaO$_{1-x}$F$_x$BiSe$_2$ and LaO$_{1-x}$F$_x$BiS$_2$ crystals have been compressed simultaneously up to 9 GPa with a typical step of 0.5 GPa (Re, $D_g$ = 300 µm, $D_c$ = 600 µm, Mar165 CCD detector). Finally, compression of LaO$_{1-x}$F$_x$BiSe$_2$ and LaO$_{1-x}$F$_x$BiS$_2$ crystals was performed simultaneously at 5 K up to 24 GPa in order to probe their superconducting regime (stainless steel gasket, $D_g$ = 300 µm, $D_c$ = 600 µm, Mar165 CCD detector). For all measurements He gas was used as a pressure transmitting medium since it preserves excellent hydrostaticity up to at least 50 GPa [26]; ruby spheres were used as pressure gauges. No bridging, i.e. sample clamping between opposite diamonds, was observed for all the pressure ramps. Single crystal diffraction data were processed with a CrysalisPro suite [27] and refined with a SHELXL program [28].

**Results and discussion**

Reciprocal spaces of the SCI phase for both LaO$_{1-x}$F$_x$BiSe$_2$ and LaO$_{1-x}$F$_x$BiS$_2$ exhibit regular features fully compatible with a *P*4/*nmm* CeBiOS$_2$-type structure (Fig. 2, left, Fig. 3, top). Neither reflections related to ferro-lattice-distortions observed for LaO$_{1-x}$F$_x$BiS$_2$ [29] nor peak splitting indicative of a monoclinic symmetry reported for LaOBiS$_2$ [30] are present in the SCI phase of LaO$_{1-x}$F$_x$BiSe$_2$ and LaO$_{1-x}$F$_x$BiS$_2$. Reported symmetry variations could originate from a difference in the exact compositions of the studied samples. Indeed, it was shown that the energy difference between various polytypes of LaOBiS$_2$ is rather small and, consequently, structural distortions can be tuned by chemical substitutions [31].

The SCI-SCII transition in the LaO$_{1-x}$F$_x$BiSe$_2$ sample occurred between 3.7 and 4.8 GPa (data for SCII taken at 5.9 GPa are shown on Fig. 2, right). The most striking difference between the SCI and SCII phases is a change in the nature of interactions along the *c* axis, direction of stacking of the BiSe$_2$ and LaO$_{1-x}$F$_x$ layers (Fig. 1). This is evident from the appearance of diffuse scattering propagating along the *c** direction concomitant with a smearing of the corresponding Bragg reflections (Fig. 2, right bottom) which, for layered systems, is indicative of a presence of stacking faults [32,33]. Similar behavior was observed in LaO$_{1-x}$F$_x$BiS$_2$ (Fig. 3, bottom). This indicates that the application of pressure induces anisotropic structural changes which results in a weakening of the interlayer interactions along the *c* axis.

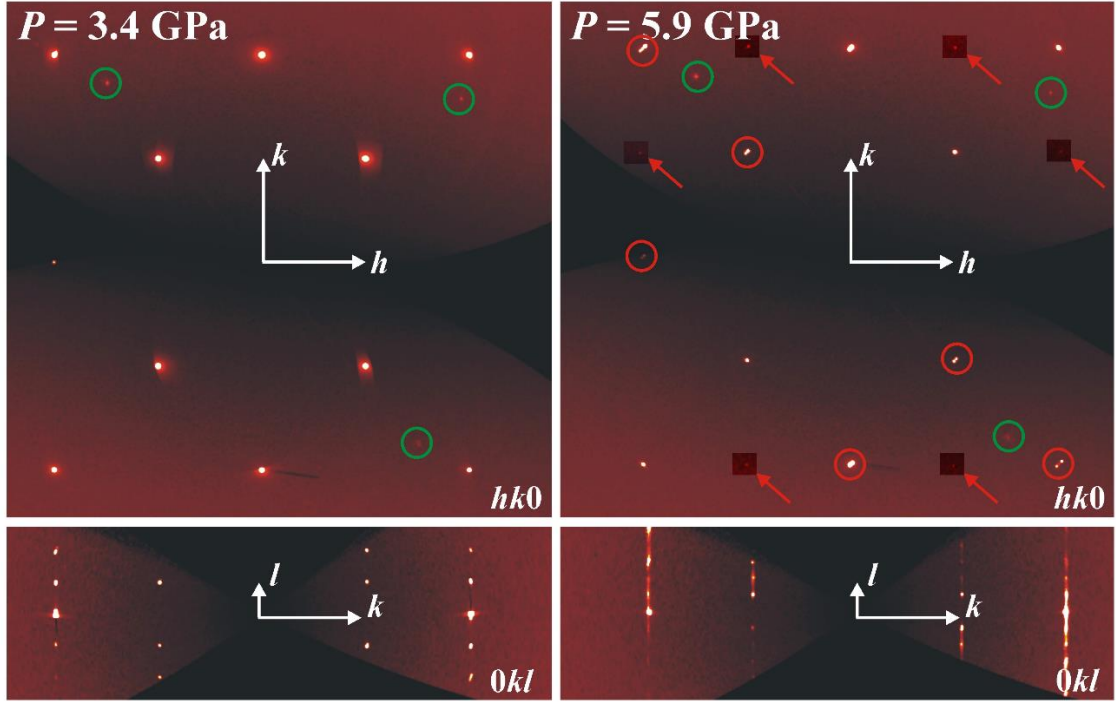

Fig. 2. Slices of the reciprocal space of LaO$_{1-x}$F$_x$BiSe$_2$ before (left, SCI phase) and after the $P$-induced transition (right, SCII phase). Reflections marked with green circles are not commensurate with the reciprocal lattice of LaO$_{1-x}$F$_x$BiSe$_2$ and originate from sample environment (diamonds, ruby spheres). Red circles (right) highlight a splitting of Bragg reflections resulting from twinning in the SCII phase induced by a tetragonal-orthorhombic symmetry lowering. Red arrows (right) indicate Bragg reflections of the HP SCII phase not present in the LP SCI phase; due to their relatively weak intensities contrast of the corresponding regions was enhanced for visualization purposes.

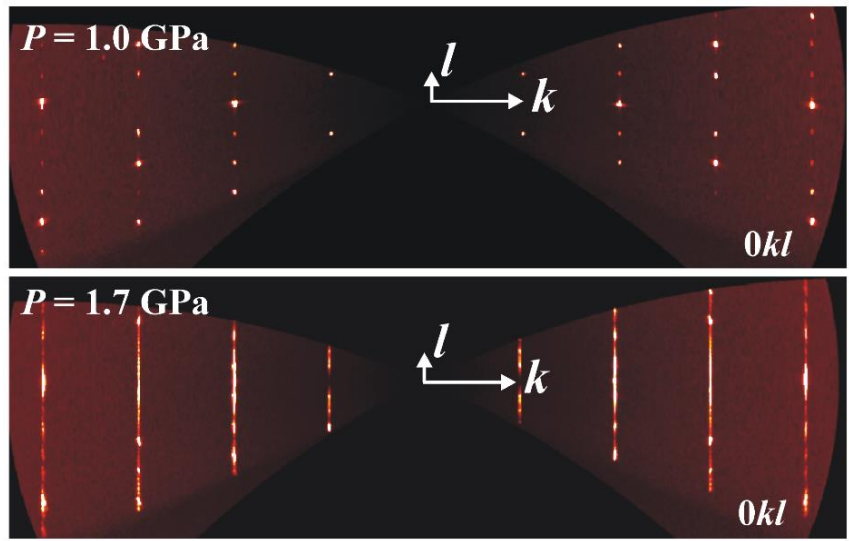

Fig. 3. Slices of the reciprocal space of LaO$_{1-x}$F$_x$BiS$_2$ before (top, SCI phase) and after the $P$-induced transition (bottom, SCII phase) showing appearance of diffuse rods in SCII.

Indeed, structural response of $LaO_{1-x}F_xBi(Se/S)_2$ to the application of HP is highly anisotropic, as concluded from single crystal diffraction data for $LaO_{1-x}F_xBiSe_2$. Firstly, while the $a$ and $c$ structural parameters of SCI $LaO_{1-x}F_xBiSe_2$ decrease upon application of pressure (Fig. 4, red and blue curves), a corresponding $a/c$ ratio increases (Fig. 4, black curve). This indicates that relatively to a monotonic decrease along the $c$ direction structure expands in the $ab$ plane.

Atomic parameters of $LaO_{1-x}F_xBiSe_2$ also exhibit a strong anisotropy upon compression. The biggest change is observed for the Bi-Se2 distances which rapidly shrink as a function of pressure (Fig. 5, red lines). The relative decrease is equal to ~12% and results in a flattening of the BiSe5 pyramids (Fig.1, middle, the shown amplitude of the flattening is exaggerated for illustration purposes). Interestingly, the La-Se2 distances which correspond to the BiSe2-LaO/F interlayer interaction increase with pressure (Fig. 5, black lines), thus these bonds exhibit negative linear compressibility. The resulting weakening in the interlayer bonding yields a strong reduction in a long range crystalline order along the $c$ axis with a subsequent phenomenon of stacking faults (Fig. 1, right) with an associated appearance of diffuse scattering in the reciprocal space. The latter is in the origin of the peak broadening observed on $P$-dependent powder diffraction data for SCII phase [19]. The BiSe2 and LaO/F layers stay rather rigid in the $ab$ plane and the corresponding bonding exhibits gradual compression with a much smaller relative changes of <3% for La-O/F (Fig. 5, green lines) and ~1% for Bi-Se1 distances (Fig. 5, blue lines). Results of structural refinement for all the datapoints presented on Figs. 4 and 5 are available in Supplemental Material [34].

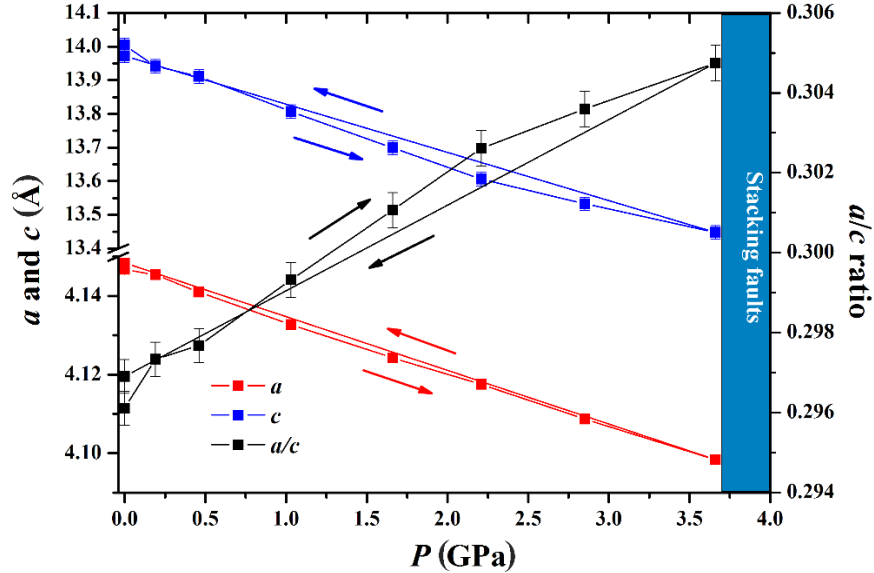

Fig. 4. $P$-dependent behavior of the $a$ and $c$ parameters of SCI $LaO_{1-x}F_xBiSe_2$ and a corresponding $a/c$ ratio. Arrows pointing to the right indicate data curves obtained on compression and the ones pointing to the left indicate decompression lines connecting maximum and ambient pressure datapoints.

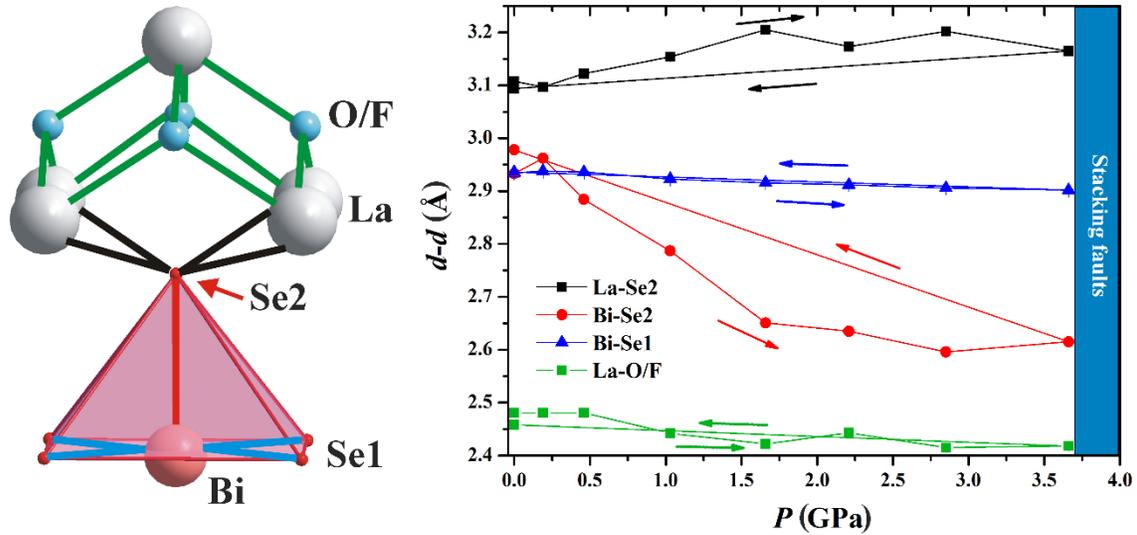

Fig. 5. Bonding evolution in LaO$_{1-x}$F$_x$BiSe$_2$ as a function of pressure. Arrows pointing to the right indicate data curves obtained on compression and the ones pointing to the left indicate decompression lines connecting maximum and ambient pressure datapoints (right figure).

Another interesting feature in the reciprocal space of SCII LaO$_{1-x}$F$_x$BiSe$_2$ is a twinning-induced splitting of Bragg reflections (Fig. 2, top right, marked with red circles). This twinning originates from the tetragonal-orthorhombic symmetry lowering associated with the SCI-SCII structural transformation and the two orthorhombic domains are rotated by ~1.8° with respect to each other. The *a* and *b* parameters of orthorhombic SCII decrease rather monotonically and equally at low pressures but feature strong divergence above 30 GPa (Fig. 6, left). Indeed, the corresponding orthorhombic distortion, defined as $\delta = (a-b)/(a+b)$, is rather small below 30 GPa ($\delta$ ~ 0.002) and reaches the value of 0.025 at the highest reached pressure of 55 GPa. This strong distortion indicates a change in the bonding interactions within the BiSe$_2$ and LaO$_{1-x}$F$_x$ layers at high pressure with a resulting large structural anisotropy in the *ab* planes. Full structural analysis for SCII LaO$_{1-x}$F$_x$BiSe$_2$ based on single crystal diffraction data, similar to the one presented above for the SCI phase, was not possible since significant part of the diffraction signal was leaked to the diffuse scattering propagating along the *c* direction and, therefore, the SCII structure cannot be treated in a classical 3D approach. Spectroscopic probes like EXAFS could be useful to track bonding evolutions within the BiSe$_2$ and LaO$_{1-x}$F$_x$ layers for the SCII phase. We stress, however, that the SCII phase remains metrically orthorhombic ($\alpha = \beta = \gamma = 90°$) up to at least 55 GPa, i.e. no symmetry lowering is observed above 30 GPa and, therefore, the monoclinic $P2_1/m$ symmetry with $\beta = 97.3°$ suggested in [19] is not compatible with our data. The latter structure was proposed based on *P*-dependent powder diffraction data which is, by definition, not resolved in reciprocal space unlike data obtained with single crystal-based techniques. In powder-based data diffuse scattering is projected along with Bragg peaks on intensity vs. angle 2D plots and is manifested as peak broadening or intensity galos. Symmetry lowering proposed in [19] was able to partially account for these effects simply by generating larger number of reflections but, as expected, the final model could not be satisfying ($R_{wp}$ = 25,5%) since this approach is experimentally limited.

Even though the diffuse scattering along the $c^*$ direction emerges by the expense of intensities of the corresponding Bragg peaks it was still possible to track positions for many of the parent peaks of LaO$_{1-x}$F$_x$BiSe$_2$ from the collected high resolution single crystal diffraction data. At the SCI-SCII transition point the $c$ parameter, which corresponds to the direction of the stacking, features a decrease of 5.5% with a resulting reduction by 5% in the unit-cell volume (Fig. 6, right). These abrupt changes in structural parameters are characteristic of first-order structural transition. Experimental equations of states (EOS) for the SCI and SCII phases were obtained by fitting the corresponding $V$-$P$ dependences with Murnaghan (Eq. 1) and 3$^{rd}$ order Birch-Murnaghan (Eq. 2) equations, correspondingly (Table 1). Here $V_0$ - zero pressure volumes, $B_0$ - bulk moduli and $B_0'$ - their first pressure derivatives. The SCII is less compressible than the parent low pressure SCI phase and the corresponding zero pressure volume, $V_0$, is diminished by ~4%.

$$V(P) = V_0 (1 + B_0' \frac{P}{B_0})^{-\frac{1}{B_0'}} \quad \text{Eq. 1}$$

$$P(V) = \frac{3B_0}{2}\left[(\frac{V_0}{V})^{\frac{7}{3}} - (\frac{V_0}{V})^{\frac{5}{3}}\right]\left\{1 + \frac{3}{4}(B_0' - 4)\left[(\frac{V_0}{V})^{\frac{2}{3}} - 1\right]\right\} \quad \text{Eq. 2}$$

Table 1. Coefficients of the Murnaghan and 3$^{rd}$ order Birch-Murnaghan EOS for the SCI and SCII phases of LaO$_{1-x}$F$_x$BiSe$_2$, respectively.

| Phase | $V_0$, Å$^3$ | $B_0$, GPa | $B_0'$, GPa |
|---|---|---|---|
| LaO$_{1-x}$F$_x$BiSe$_2$ SCI | 241.5(3) | 44(5) | 8(3) |
| LaO$_{1-x}$F$_x$BiSe$_2$ SCII | 231.3(7) | 67(3) | 4.3(2) |

Another difference between the SCII and SCI phases, in addition to the presence of stacking faults and the symmetry lowering, is the presence of new weak reflections with odd $h+k$ indexes which violate the $n$ glide plane of the parent $P4/nmm$ symmetry (Fig. 2, top right, marked with red arrows). The actual space group cannot be assigned for the SCII structure due to a lack of long-range correlations along the $c$ stacking direction. The new $hk0$ pattern is, however, compatible with a model proposed in [29] featuring ferro-lattice-distortions.

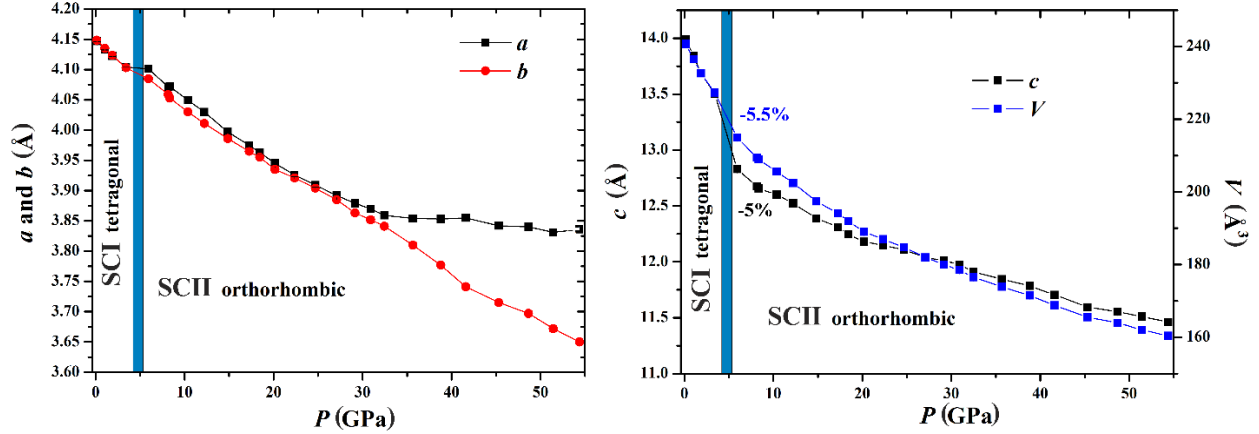

Fig. 6. *P*-dependent behavior of the *a*, *b* (left), *c* (right) structural parameters and unit-cell volume (right) of $LaO_{1-x}F_xBiSe_2$ up to *P* = 55 GPa.

Although structure of the diffuse rods of $LaO_{1-x}F_xBiSe_2$ is not regular (Fig. 2, bottom right), new intensity maximums with $q = 0.5c*$ are clearly visible and some of those have even higher intensities than the parent $q = 1c*$ reflections. This indicates that in addition to original ($q = 1c*$) correlations of the parent SCI phase new correlations, which correspond to the doubling of the *c* parameter of the SCI phase ($q = 0.5c*$), emerge. A corresponding stacking model could comprise alternative rotations/shifts of the $BiS_2$ bilayers in opposite directions with respect to the reference LaO/F framework. Interestingly, the $LaO_{1-x}F_xBiS_2$ phase features much more homogenous structure of the diffuse rods (Fig. 3, bottom) indicating that the corresponding correlations are mostly of a short range order. In addition, the SCI-SCII transition for the $BiS_2$-based phase occurs between 1.0 and 1.7 GPa, which is lower than the corresponding transition pressure in $LaO_{1-x}F_xBiSe_2$. Therefore the interlayer $BiS_2$ – LaO/F bonding is weaker compared to the corresponding interactions in the $BiSe_2$-based phase.

At ambient temperature the SCI-SCII transition is reversible, i.e. diffuse rods disappear upon decompression. Nevertheless, the HP annealing likely stabilizes the SCII phase which is manifested through a broadening of reflections on experimental powder diffraction data [21-25]. Single crystal diffraction experiments at HP-HT conditions are needed to confirm this likely scenario. Finally, compression at 5 K was performed to study phase behavior for both $LaO_{1-x}F_xBiSe_2$ and $LaO_{1-x}F_xBiS_2$ at superconducting conditions. At 5 K the SCI-SCII transition in $LaO_{1-x}F_xBiSe_2$ occurred between 2.9 and 5.2 GPa, i.e. in the same pressure range as at ambient temperature. Similarly, the $LaO_{1-x}F_xBiS_2$ sample transited already at 1.7 GPa (the first pressure point collected at 5 K). Thus the SCI-SCII phase boundary in a *T-P* space is rather vertical for both $LaO_{1-x}F_xBiSe_2$ and $LaO_{1-x}F_xBiS_2$ and the corresponding phase diagram is shown in Fig. 7.

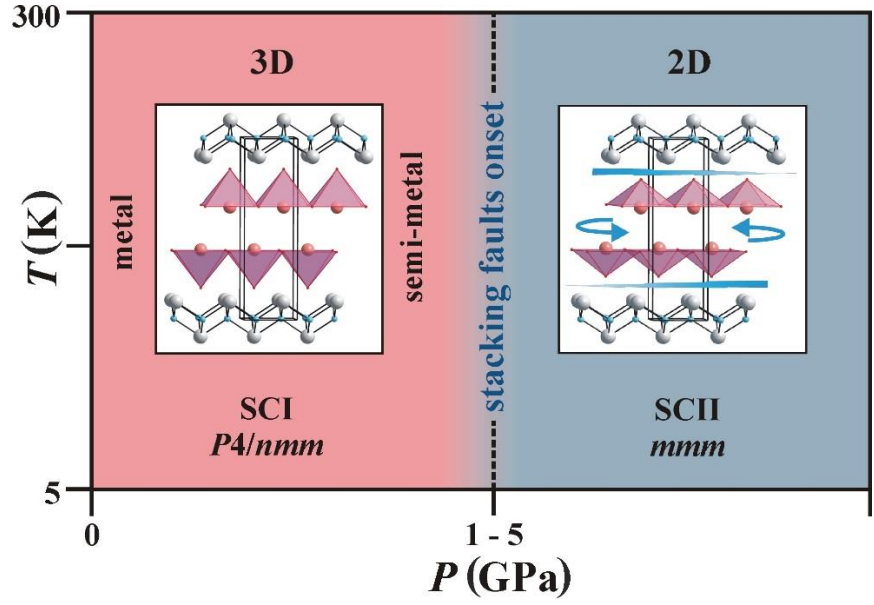

Fig. 7. Schematic $T$-$P$ phase diagram of LaO$_{1-x}$F$_x$BiSe$_2$ and LaO$_{1-x}$F$_x$BiS$_2$. The Laue *mmm* class is indicated for the orthorhombic SCII phase. Conduction properties refer to the LaO$_{1-x}$F$_x$BiSe$_2$ phase and have been obtained from *ab initio* calculations based on experimental $P$-dependent structural data for LaO$_{1-x}$F$_x$BiSe$_2$ (Supplemental Material [34]).

Electronic density of states (DOS) have been calculated for LaO$_{1-x}$F$_x$BiSe$_2$ structures at 0 and 3.7 GPa, right before the onset of the SCI-SCII transition, in order to deduce the origin of the decrease in $T_c$ observed experimentally in this phase as a function of pressure [5,6]. Calculations were performed with a Wien2k 18.2 code [35] using the PBE-GGA exchange correlation potential [36]. The Brillouin zone (BZ) was divided with a k-mesh of 784 points (56 k-points in the irreducible region of the BZ). For each pressure structural parameters obtained from the corresponding diffraction data were used; O/F atomic position was considered as fully occupied solely by oxygen atoms ($x = 0$).

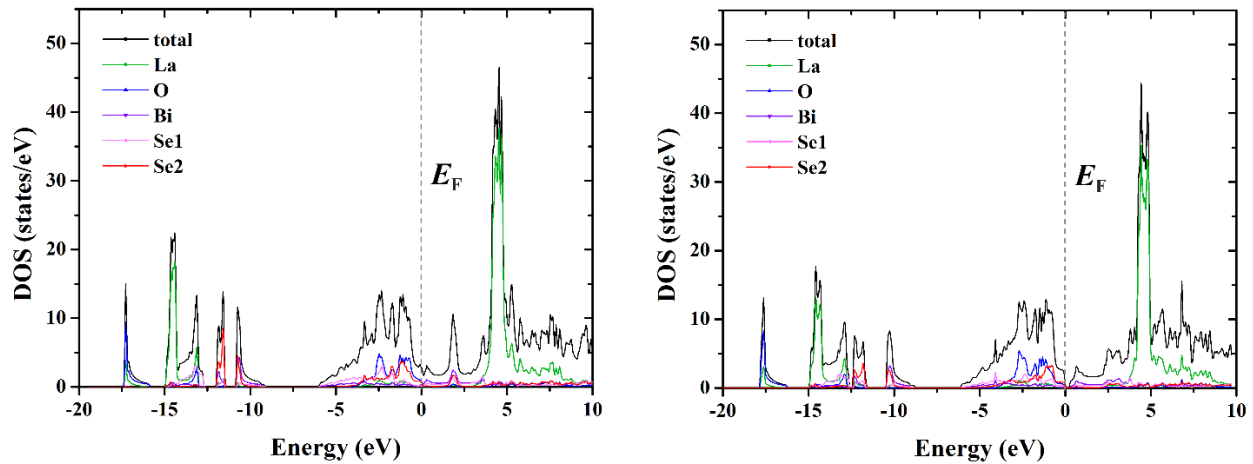

Fig. 8. Calculated DOS for the LaO$_{1-x}$F$_x$BiSe$_2$ structures at 0 (left) and 3.7 GPa (right).

At ambient pressure the electronic states of LaO$_{1-x}$F$_x$BiSe$_2$ ($x = 0$) at the Fermi level, $E_F$, are populated essentially by $p$-levels of Se2 and O atoms (Fig. 8, left, red and blue curves) indicating a presence of an inter-layer charge transfer, similarly to the LaO$_{1-x}$F$_x$BiS$_2$ system [37]. Electron doping by partial substitution of F for O directly replenishes electronic states at $E_F$ therefore enhancing metallic properties of the parent LaOBiSe$_2$ phase. Contrary, application of HP nearly depletes electronic states at $E_F$ with a complete depletion right above $E_F$ producing a small bandgap of 0.04 eV (Fig. 8, right). The resulting semi-metallic behavior is associated with a decrease in $T_c$ for this phase. However, the following SCI-SCII transition enhances two-dimensionality in LaO$_{1-x}$F$_x$BiSe$_2$, as demonstrated above, resulting in a phase with superior superconducting properties.

**Conclusions**

HP strongly enhances two-dimensionality in LaO$_{1-x}$F$_x$Bi(Se/S)$_2$. The associated reduction in long-range order along the direction of stacking of Bi(Se/S)$_2$ and LaO/F layers induces formation of stacking faults. The HP SCII LaO$_{1-x}$F$_x$BiSe$_2$ phase better preserves long-range order of the parent SCI phase compared to the S-containing one. This is evident from a more homogeneous structure of diffuse rods for LaO$_{1-x}$F$_x$BiS$_2$ and, therefore, this system is characterized by weaker interlayer interactions which results in lower SCI-SCII transition pressure. Studied LaO$_{1-x}$F$_x$Bi(Se/S)$_2$ systems illustrate how structural two-dimensionality is critical for layered superconductors. Targeted enhancement of 2D-structure of layered systems either by external (pressure, temperature) or internal (chemical doping, substitutions) means should be considered as another pathway to further improve their superconducting performance.

**Acknowledgments**

Authors acknowledge the European Synchrotron Radiation Facility for the allocation of in-house beamtime and financial support. This work was also partly supported by statutory budget of Faculty of Chemistry, Warsaw University of Technology.